\documentclass[printer]{aa}

\usepackage{graphicx}
\usepackage{txfonts}
\usepackage[draft]{hyperref}

\begin{document} 

\title{Timing of PSR~J2055+3829, an eclipsing black widow pulsar discovered with the Nan\c{c}ay Radio Telescope}

\author{
L.~Guillemot\inst{1,2} 
\and F.~Octau\inst{1,2}
\and I.~Cognard\inst{1,2} 
\and G.~Desvignes\inst{3}
\and P.~C.~C.~Freire\inst{3}
\and D.~A.~Smith\inst{4}
\and G.~Theureau\inst{1,2,5}
\and T.~H.~Burnett\inst{6}
}

\institute{
Laboratoire de Physique et Chimie de l'Environnement et de l'Espace -- Universit\'e d'Orl\'eans / CNRS, F-45071 Orl\'eans Cedex 02, France\\
\email{lucas.guillemot@cnrs-orleans.fr}
\and
Station de radioastronomie de Nan\c{c}ay, Observatoire de Paris, CNRS/INSU, F-18330 Nan\c{c}ay, France
\and
Max-Planck-Institut f\"ur Radioastronomie, Auf dem H\"ugel 69, D-53121 Bonn, Germany\\
\and 
Centre d'\'Etudes Nucl\'eaires de Bordeaux Gradignan, IN2P3/CNRS, Universit\'e Bordeaux 1, BP120, F-33175 Gradignan Cedex, France\\
\and
LUTH, Observatoire de Paris, PSL Research University, CNRS, Universit\'e Paris Diderot, Sorbonne Paris Cit\'e, F-92195 Meudon, France\\
\and
Department of Physics, University of Washington, Seattle, WA 98195-1560, USA\\
}

\date{Received September 15, 1996; accepted March 16, 1997}


\abstract{
We report on the timing observations of the millisecond pulsar PSR~J2055+3829 originally discovered as part of the SPAN512 survey conducted with the Nan\c{c}ay Radio Telescope. The pulsar has a rotational period of 2.089 ms, and is in a tight 3.1 hr orbit around a very low mass ($0.023 \leq m_c \lesssim 0.053$ M$_\odot$, 90\% c.l.) companion. Our 1.4 GHz observations reveal the presence of eclipses of the pulsar's radio signal caused by the outflow of material from the companion, for a few minutes around superior conjunction of the pulsar. The very low companion mass, the observation of radio eclipses, and the detection of time variations of the orbital period establish PSR~J2055+3829 as a `black widow' (BW) pulsar. Inspection of the radio signal from the pulsar during ingress and egress phases shows that the eclipses in PSR~J2055+3829 are asymmetric and variable, as is commonly observed in other similar systems. More generally, the orbital properties of the new pulsar are found to be very similar to those of other known eclipsing BW pulsars. No gamma-ray source is detected at the location of the pulsar in recent \textit{Fermi}-LAT source catalogs. We used the timing ephemeris to search ten years of \textit{Fermi} Large Area Telescope (LAT) data for gamma-ray pulsations, but were unable to detect any. This non-detection could be a consequence of the pulsar's large distance compared to those of known gamma-ray millisecond pulsars outside of globular clusters. We finally compared the mass functions of eclipsing and non-eclipsing BW pulsars and confirmed previous findings that eclipsing BWs have higher mass functions than their non-eclipsing counterparts. Larger inclinations could explain the higher mass functions of eclipsing BWs. On the other hand, the mass function distributions of Galactic disk and globular cluster BWs appear to be consistent, suggesting, despite the very different environments, the existence of common mechanisms taking place in the last stages of evolution of BWs.
}


\maketitle

\section{Introduction}

Millisecond pulsars (MSPs) are rotation-powered neutron stars with very short rotational periods ($P \lesssim 30$ ms), which are believed to have been spun-up by the accretion of matter and thus the transfer of angular momentum from a binary companion \citep{Bisnovatyi1974,Alpar1982}. While most known MSPs reside in binary systems around He or CO-white dwarf companions, a fraction of these objects are in tight binaries (with orbital periods $P_b \lesssim 1$ day) with very light companions ($m_c \lesssim 0.05$ M$_\odot$). In these `black widow' (BW) systems, the tidally-locked companion star is irradiated by the pulsar's intense plasma wind, and outflowing material from the companion causes long eclipses of the pulsar's radio pulses (of $\sim$ 5--10\% of the orbital period, typically), which are observed when the companion passes between the observer and the pulsar, provided the orbital inclination of the binary system is high enough. The exact mechanism under which the radio emission from the pulsar is eclipsed is currently unknown \citep[see discussions in][]{Thompson1994,Wadiasingh2017,Polzin2018}. However, it is clear that eclipsing MSPs are invaluable laboratories of plasma physics \citep[see e.g.][for a recent example]{Main2018}, and BW systems may also represent a formation channel for isolated MSPs \citep{Ruderman1989}. Nevertheless, although three decades have passed since the discovery of the original BW pulsar B1957+20 \citep{Fruchter1988}, and despite the recent discoveries of many BW pulsars in \textit{Fermi}-LAT unassociated sources through radio observations \citep[e.g.,][]{Ray2012,Bhattacharyya2013,Cromartie2016}, the known population of BW pulsars observed to exhibit radio eclipses remains small (see Section~\ref{s:massfuncs}). Discovering and studying more of these eclipsing systems is thus of prime importance.

The Nan\c{c}ay Radio Telescope (NRT) is a meridian telescope equivalent to a 94-m parabolic dish located near Orl\'eans (France). Owing to its design, the NRT can track objects with declinations $\delta > -39^\circ$ for approximately one hour around culmination, and is thus well suited for the long-term timing of pulsars, e.g. for the study of individual objects \citep[see e.g.][for recent examples]{Cognard2017,Octau2018} or for searching low-frequency gravitational waves from supermassive black hole binaries, using Pulsar Timing Arrays \citep[PTAs, see e.g.][]{Desvignes2016}. With the goal of identifying new exotic pulsar systems or highly-stable MSPs suitable for PTA studies, the SPAN512 pulsar survey  \citep{Desvignes2013,Octau2016,Desvignes2019} was conducted between 2012 and 2018 at the NRT. As part of this survey, new pulsars were searched for at intermediate Galactic latitudes ($3.5\degr < |b| < 5\degr$) and away from the inner Galaxy (Galactic longitudes $74\degr < l < 150\degr$). Observations were conducted at 1.4~GHz with 0.5~MHz frequency channels over a total bandwidth of 512~MHz and a fine time resolution of 64~$\mu$s, to be sensitive to faint and distant MSPs. The data were searched for pulsars with dispersion measures (DMs) up to 1800 pc cm$^{-3}$, with a moderate acceleration search in the Fourier domain (the \texttt{zmax} parameter was set to 50 in PRESTO analyses) to be sensitive to pulsars in binary systems. Searches for periodic signals in the data from this survey so far led to the discovery of one ``ordinary'' (\textit{i.e.}, non-millisecond) pulsar, PSR~J2048+49, and two MSPs, PSRs~J2055+3829 and J2205+6012. Details on the survey, the data analysis, and the discovered pulsars will be reported in \citet{Desvignes2019}. In the present paper we report on the results from the timing of PSR~J2055+3829, an MSP in an eclipsing BW system, and from the analysis of the pulsar's radio eclipses. In Section~\ref{s:analysis} we describe the radio timing observations and the results from the analysis of the timing data. In Section~\ref{s:eclipses} we present observations of eclipses of PSR~J2055+3829 at 1.4~GHz, and analyses of the data taken around superior conjunction of the pulsar. In the following section (Section~\ref{s:massfuncs}), we present comparisons of the mass function distributions for eclipsing and non-eclipsing BWs, and for Galactic disk and globular cluster BWs. Finally, Section~\ref{s:summary} summarizes our findings.

\section{Observations and data analysis}
\label{s:analysis}

\subsection{Radio observations and timing analysis}
\label{s:timing}

The discovery of PSR~J2055+3829 was described in detail by \citet{Octau2017}. In summary, the analysis of an 18-min observation conducted as part of the SPAN512 survey on 5 December 2013 (MJD 56631) at 1.4~GHz with PRESTO pulsar searching routines \citep{PRESTO} revealed a candidate pulsar with a period of 2.08 ms and a DM of 91.9 pc cm$^{-3}$. Confirmation observations of the same sky direction (R.A. $=$ 20:55:04, decl. $=$ $+$38:37:23) were carried out with the NRT on 19 October 2015 (MJD 57314) for 14 minutes and on 20 October 2015 (MJD 57315) for 67 minutes, both of them resulting in firm detections of the pulsar, with a statistical significance well above 10$\sigma$. After the first few days of timing of the MSP, it became clear that it is in a compact, low eccentricity orbit around a very low mass companion, with an orbital period of 3.1 hr.

\begin{figure}
\centering
\includegraphics[width=0.95\columnwidth]{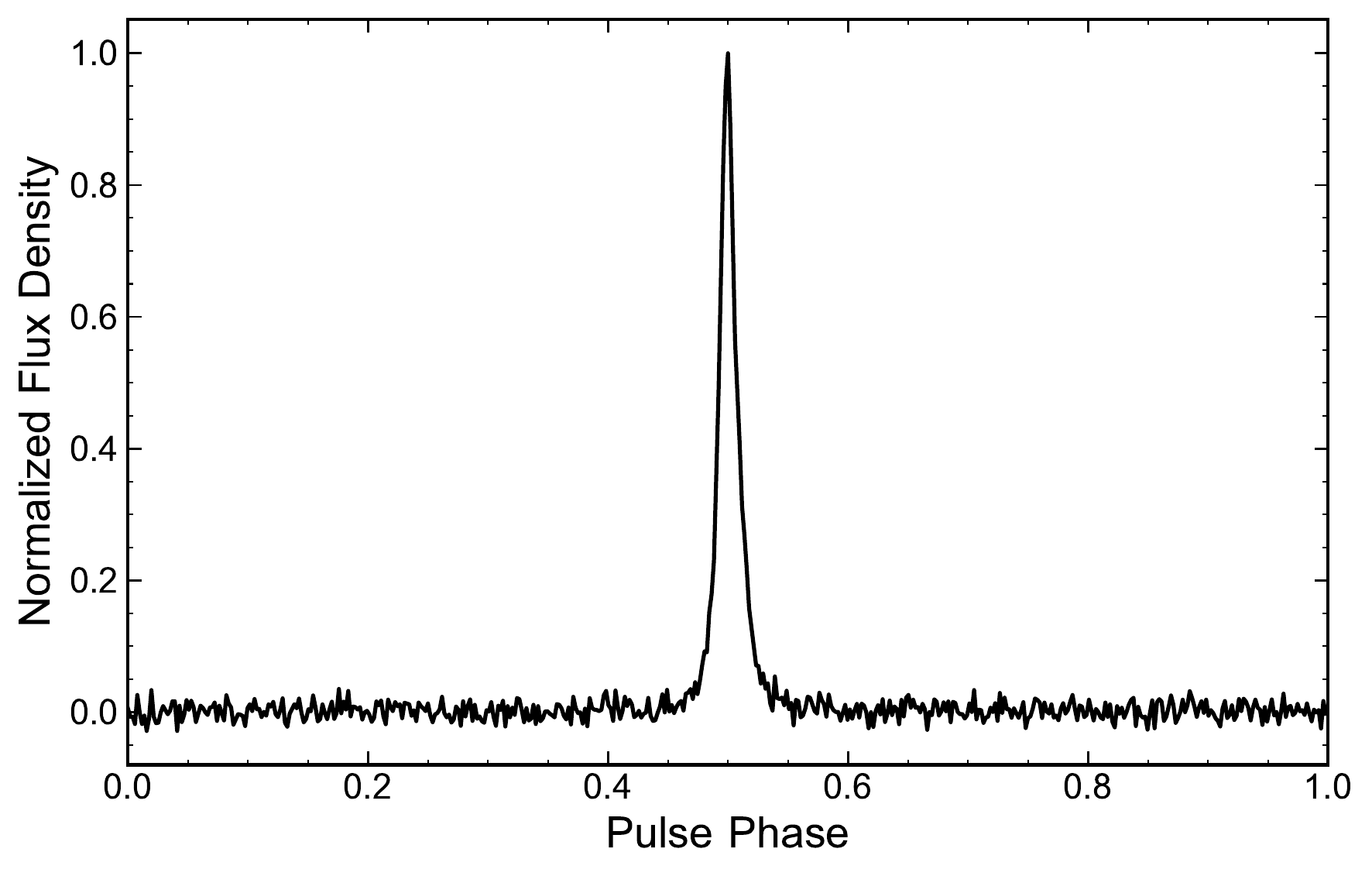}
\caption{Pulse profile for PSR~J2055+3829 at 1.4~GHz, formed by integrating 48.8 hr of coherently dedispersed observations with the Nan\c{c}ay Radio Telescope and the NUPPI backend. Profiles recorded at orbital phases between 0.1 and 0.4 were excluded from the summation, to avoid contamination of the integrated profile from dispersive delays caused by ionized material near the companion star. To improve readability we divided the number of phase bins by four, to form a profile with 512 phase bins.}
\label{profile}
\end{figure}

At that stage, our initial timing solution for PSR~J2055+3829 enabled us to time the pulsar accurately over the typical duration of NRT pulsar observations of 1 hour. We thus commenced regular timing observations of the new MSP. The bulk of the observations were made at 1.4 GHz (with an exact central frequency of 1484 MHz), and a few observations at higher frequencies of 2154 MHz or 2539 MHz were also conducted. The latter observations resulted in much weaker detections of the pulsar; hence, most of the observations presented here were done at 1.4 GHz. We used the NUPPI pulsar observation backend, a version of the Green Bank Ultimate Pulsar Processing Instrument\footnote{\url{https://safe.nrao.edu/wiki/bin/view/CICADA/NGNPP}} designed for Nan\c{c}ay \citep[see][for a description]{Cognard2013}. In these timing observations, 128 channels of 4~MHz each are coherently dedispersed in real time and the time series are folded online at the expected topocentric pulsar period. We used the PSRCHIVE software library \citep{PSRCHIVE} to clean the data of radio frequency interference (RFI) and used the \textsc{SingleAxis} method of PSRCHIVE to calibrate the polarization information. High signal-to-noise (S/N) NUPPI observations of PSR~J2055+3829 made in timing mode at 1.4~GHz were combined to form an average pulse profile, shown in Figure~\ref{profile}. As will be discussed later, the pulsar exhibits eclipses around superior conjunction (defined as orbital phase 0.25) and extra delays in the radio pulses recorded during eclipse ingress and egress caused by ionized material near the companion star are measured. We therefore conservatively excluded data at orbital phases between 0.1 and 0.4 when forming the average pulse profile. A total of 48.8~hr of coherently dedispersed data were summed in the process. As can be seen from the figure, the pulsar's pulse profile at 1.4~GHz consists of a single, slightly asymmetric sharp peak, with no evidence for a secondary emission component. For the observations summed in the process we measure an average radio flux density at 1.4 GHz of $S_{1400} = 0.10 \pm 0.04$~mJy, where the uncertainty reflects the standard deviation of the individual flux density measurements. We note that since a large number of individual flux density measurements were used in the calculation, the average flux density quoted above is unlikely to be biased by scintillation effects.

A high S/N reference profile for PSR~J2055+3829 at 1.4~GHz was generated by smoothing the integrated pulse profile shown in Figure~\ref{profile}, and times of arrival (TOAs) were generated by cross-correlating the reference profile with the individual profiles. This cross-correlation was performed using the ``Fourier domain with Markov chain Monte Carlo algorithm'' implemented in the \texttt{pat} routine of PSRCHIVE, which properly estimates TOA uncertainties in the low S/N regime. For each observation, we generated one TOA per 10 min and per 128 MHz of bandwidth, so that each TOA covers less than 10\% of an orbit, and in order to track potential time variations of the DM. We analyzed the TOA data using the \textsc{Tempo2} pulsar timing package \citep{TEMPO2}. The measured topocentric TOAs were converted to Coordinated Barycentric Time (TCB) using the DE436 Solar system ephemeris\footnote{\url{https://naif.jpl.nasa.gov/pub/naif/JUNO/kernels/spk/de436s.bsp.lbl}}, and accounting for the known clock corrections for the NUPPI backend. For the timing analysis, we again discarded TOAs corresponding to orbital phases between 0.1 and 0.4.

\begin{figure*}
\centering
\includegraphics[width=16cm]{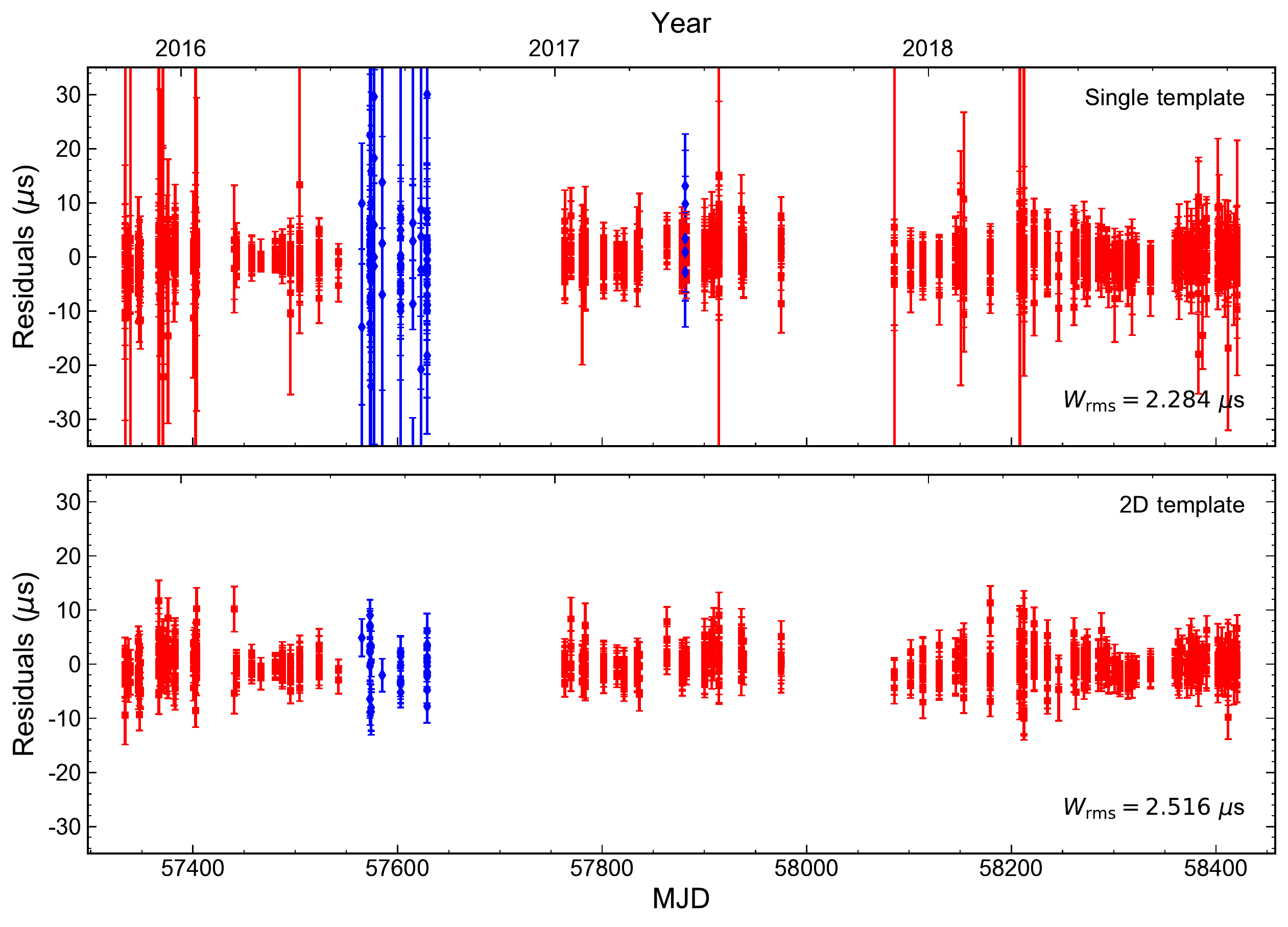}
\caption{Post-fit timing residuals as a function of time for the best-fit timing solutions presented in Table~\ref{ephem}, considering non-eclipsing binary phases only. Residuals shown in the upper panel are based on TOAs extracted by using a single template profile representative of the full 512-MHz bandwidth at 1.4 GHz, while those in the lower panel are based on TOAs obtained by fitting a template profile with the full frequency resolution to the data. Details on the TOA integration times, frequency bandwidths, and extraction procedures are given in Section~\ref{s:analysis} for the upper panel, and in Section~\ref{s:eclipses} for the lower panel. Residuals shown as red squares (resp., blue diamonds) correspond to observations made at 1.4 GHz (resp., 2.1 and 2.5 GHz).
In the upper panel (respectively, the lower panel), TOA uncertainties were multiplied by a correction factor (EFAC, see Section~\ref{s:timing}), of 1.074 (resp., 1.726).
}
\label{res_time}
\end{figure*}

The best-fit timing solution which minimizes the differences between measured TOAs and those predicted by \textsc{Tempo2} (the so-called ``timing residuals'') is presented in Table~\ref{ephem} (see results in the `Single template' column), and the timing residuals as a function of time are displayed in Figure~\ref{res_time}. PSR~J2055+3829 being in a low-eccentricity orbit, we used the ELL1 binary timing model \citep{Lange2001} which is well-suited for the fitting of orbital parameters in this configuration. In addition to the binary parameters, we fit for the pulsar's sky position, proper motion, spin frequency and its first time derivative, and the DM and its first time derivative. To account for potential biases in the determination of the TOA uncertainties, we used the EFAC and EQUAD factors of \textsc{Tempo2}. The EQUAD parameter was found to be negligible; we thus set it to 0 in the final fit. On the other hand, an EFAC very close to 1 was found, indicating realistic TOA uncertainties. As can be seen from Table~\ref{ephem}, the best-fit Laplace-Lagrange parameters $\eta$ and $\kappa$ are only marginally significant. A fit with $\eta = \kappa = 0$ (\textit{i.e.}, a circular orbit) leads to slightly higher $\chi^2$ and RMS residual values, and best-fit parameters consistent with those listed in Table~\ref{ephem}.

\begin{table*}
\caption[]{Parameters for PSR~J2055+3829 derived from the analysis of the NRT timing data using \textsc{Tempo2}. Numbers in parentheses are the nominal 1$\sigma$ statistical uncertainties in the last digits quoted. The reference epoch of the pulsar's spin, astrometric and DM parameters is MJD 57900, and was chosen to be close to the center of the timing dataset. Epochs are given in Barycentric Coordinate Time (TCB), and the DE436 planetary ephemeris was used. The DM-derived distance was estimated using the YMW16 model of free electron density \citep{YMW16}. We assumed a pulsar mass of 1.4~M$_\odot$ for the calculation of minimal companion masses, and a moment of inertia $I$ of $10^{45}$ g cm$^2$ for the calculation of $\dot E$, $B_s$ and $B_\mathrm{LC}$.}
\label{ephem}
\centering
\begin{tabular}{lcc}

\hline
\hline
\multicolumn{3}{c}{Data reduction parameters} \\
\hline
Analysis type \dotfill & Single template & 2D template \\
Span of timing data (MJD) \dotfill & $57333$ -- $58421$ & $57333$ -- 58421 \\
Number of TOAs \dotfill & $1723$ & $835$ \\
Weighted RMS residual ($\mu$s) \dotfill & $2.284$ & $2.516$ \\
EFAC \dotfill & $1.074$ & $1.726$ \\
EQUAD ($\mu$s) \dotfill & $0.000$ & $0.000$ \\
Reduced $\chi^2$ \dotfill & $1.000$ & $1.000$ \\
\hline
\multicolumn{3}{c}{Astrometric and spin parameters} \\
\hline
Right ascension, $\alpha$ (J2000) \dotfill & 20:55:$10.306550(4)$ & 20:55:$10.306556(4)$ \\
Declination, $\delta$ (J2000) \dotfill & +38:29:$30.90571(6)$ & +38:29:$30.90571(6)$ \\
Proper motion in $\alpha$, $\mu_\alpha \cos(\delta)$ (mas yr$^{-1}$) \dotfill & $5.92(3)$ & $5.87(5)$ \\
Proper motion in $\delta$, $\mu_\delta$ (mas yr$^{-1}$) \dotfill & $0.79(7)$ & $0.9(1)$ \\
Spin frequency, $\nu$ (Hz) \dotfill & $478.631427595910(5)$ & $478.63142759590(2)$ \\
Spin frequency derivative, $\dot \nu$ ($10^{-16}$ Hz s$^{-1}$) \dotfill & $-2.290(1)$ & $-2.289(2)$ \\
Dispersion measure, DM (pc cm$^{-3}$) \dotfill & $91.8295(2)$ & $91.8295(7)$ \\
Dispersion measure derivative, DM1 (pc cm$^{-3}$ yr$^{-1}$) \dotfill & $-0.0047(2)$ & $-0.0049(7)$ \\
Reference Epoch (MJD) \dotfill & $57900$ & $57900$ \\
\hline
\multicolumn{3}{c}{Binary parameters} \\
\hline
Orbital period, $P_b$ (days) \dotfill & $0.12959037294(1)$ & $0.12959037293(2)$ \\
Projected semimajor axis of the pulsar orbit, $x$ (lt-s) \dotfill & $0.0452618(2)$ & $0.0452616(3)$ \\
Epoch of ascending node, $T_\mathrm{asc}$ \dotfill & $57900.06984171(6)$ & $57900.06984175(9)$ \\
$\eta \equiv e \sin{\omega}$ ($10^{-5}$) \dotfill & $0.9(6)$ & $0.7(10)$ \\
$\kappa \equiv e \cos{\omega}$ ($10^{-5}$) \dotfill & $0.5(4)$ & $1.3(6)$ \\
Orbital period derivative, $\dot P_b$ ($10^{-12}$) \dotfill & $-2.00(9)$ & $-2.0(1)$ \\
\hline
\multicolumn{3}{c}{Derived parameters} \\
\hline
Orbital eccentricity, $e$ ($10^{-5}$) \dotfill & $1.0(6)$ & $1.4(7)$ \\
Mass function, $f$ ($10^{-6}$ M$_\odot$) \dotfill & $5.92832(7)$ & $5.9283(1)$ \\
Minimum companion mass, $m_\mathrm{c,min}$ (M$_\odot$) \dotfill & $0.02290$ & $0.02290$ \\
Total proper motion, $\mu_T$ (mas yr$^{-1}$) \dotfill & $5.97(4)$ & $5.93(6)$ \\
Galactic longitude, $l$ ($\degr$) \dotfill & \multicolumn{2}{c}{$80.615$} \\
Galactic latitude, $b$ ($\degr$) \dotfill & \multicolumn{2}{c}{$-4.259$} \\
DM-derived distance, $d$ (kpc) \dotfill & \multicolumn{2}{c}{$4.6(9)$} \\
Transverse velocity, $v_T$ (km s$^{-1}$) \dotfill & \multicolumn{2}{c}{$13(3)$} \\
Spin period, $P$ (ms) \dotfill & \multicolumn{2}{c}{$2.08929030219107(2)$} \\
Spin period derivative, $\dot P$ ($10^{-22}$ s s$^{-1}$) \dotfill & \multicolumn{2}{c}{$9.996(5)$} \\
Intrinsic spin period derivative, $\dot P_\mathrm{int}$ (10$^{-22}$ s s$^{-1}$) \dotfill & \multicolumn{2}{c}{$8(2)$} \\
Spin-down power, $\dot E$ ($10^{33}$ erg s$^{-1}$) \dotfill & \multicolumn{2}{c}{$3.6(7)$} \\
Surface magnetic field intensity, $B_s$ ($10^7$ G) \dotfill & \multicolumn{2}{c}{$4.2(4)$} \\
Magnetic field intensity at light cylinder, $B_\mathrm{LC}$ ($10^4$ G) \dotfill & \multicolumn{2}{c}{$4.3(4)$} \\
\hline

\end{tabular}
\end{table*}

\begin{figure}
\centering
\includegraphics[width=0.95\columnwidth]{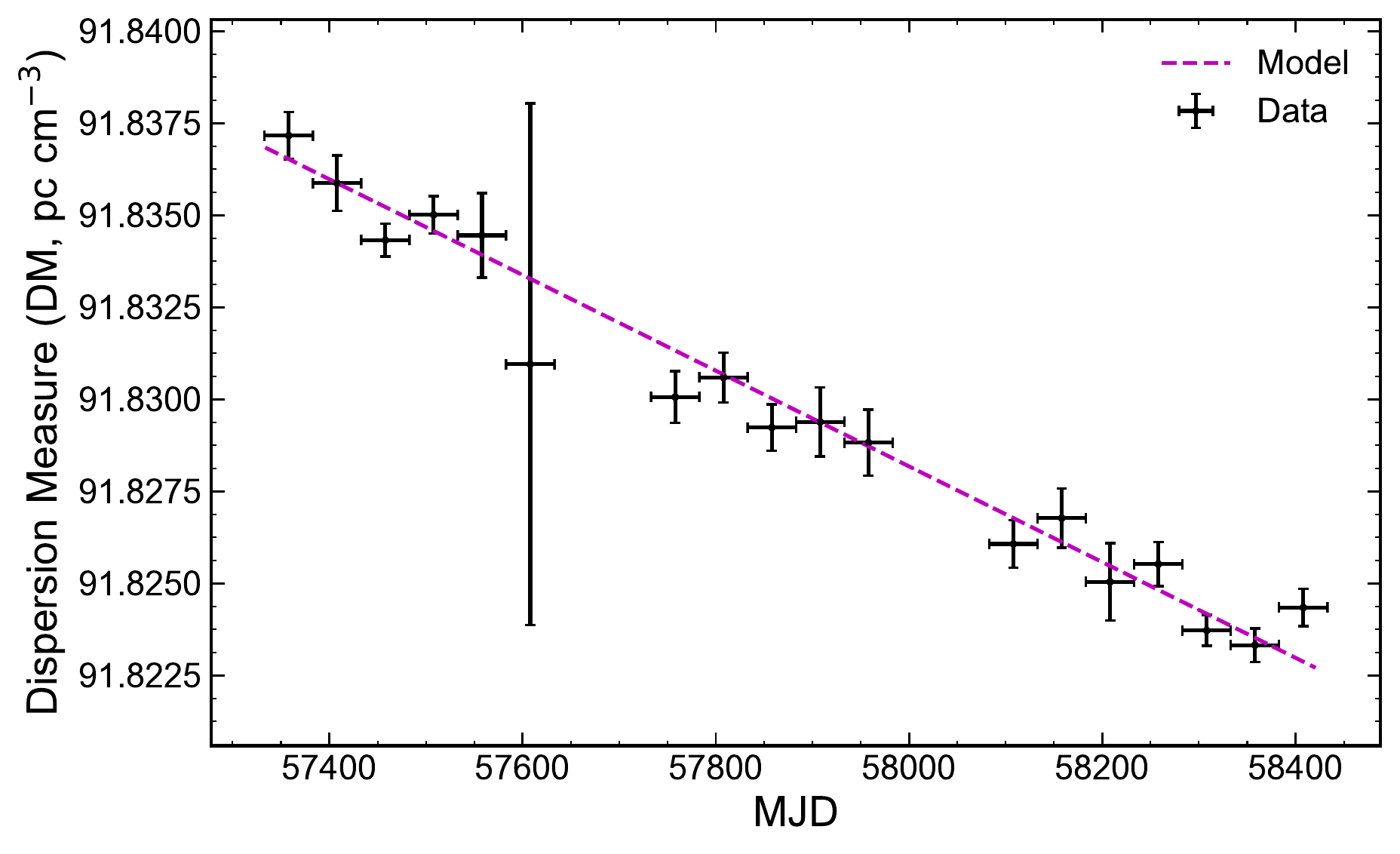}
\caption{Evolution of the Dispersion Measure (DM) of PSR~J2055+3829 as a function of time. Data points shown in black were determined by fitting constant DM values to 50-day segments of NUPPI timing data. NUPPI observations were split into four sub-bands of 128 MHz each, to allow for the fit in DM. The dashed line shows the best-fit DM model, as determined in the timing analysis (see Table~\ref{ephem}, `Single template' column). The data point at MJD $\sim$57600 corresponds to a time interval where only high frequency (2.1 and 2.5 GHz) observations were available, hence the large DM uncertainty.}
\label{DM_vs_time}
\end{figure}

The DM of the pulsar is observed to vary, and we find that a simple linear model for the DM enables us to model the multi-frequency TOAs adequately. The best-fit DM function from the timing analysis is plotted as a function of time in Figure ~\ref{DM_vs_time}. Also shown in the figure are the results of a fit of constant DM values to 50-day segments of the TOAs. The constant DM values are consistent with those predicted by the simple linear model.

Given the pulsar's DM and position, the YMW16 electron-density model of \citet{YMW16} predicts a distance $d$ of $4.6 \pm 0.9$ kpc, assuming an uncertainty of 20\%. The NE2001 model \citep{NE2001} also predicts a relatively large distance of about 4.4 kpc.  At the DM-derived distance $d \sim 4.6$ kpc, the total transverse velocity of the pulsar derived from the measured proper motion is $\sim$13 km s$^{-1}$. This low value is relatively common among other recycled pulsars \citep{Hobbs2005}. We note that the low transverse velocity of PSR~J2055+3829 combined with the rate of change of the DM of $\sim$$5 \times 10^{-3}$~pc~cm$^{-3}$~yr$^{-1}$ which is relatively high for an MSP \citep[see e.g.][]{Jones2017} suggest that the pulsar could be undergoing radial motion through a region with a strong free electron density gradient. The modest transverse proper motion leads to a small Shklovskii correction \citep{Shklovskii1970} to the apparent spin period derivative, $\dot P$, calculated as:
 
\begin{eqnarray}
\dot P_\mathrm{Shk} \simeq 2.43 \times 10^{-21} \left( \dfrac{\mu_\perp}{\mathrm{mas\ yr}^{-1}} \right) \left( \dfrac{d}{1\ \mathrm{kpc}} \right) \left( \dfrac{P}{\mathrm{s}} \right), 
\end{eqnarray}

where $\mu_\perp$ is the total transverse proper motion of the pulsar. We accounted for this kinematic effect and for the correction caused by the difference in Galactic accelerations of the pulsar and of the solar system, using the model for the rotation of the Galaxy from \citet{Carlberg1987} and \citet{Kuijken1989}, to derive the intrinsic spin period derivative $\dot P_\mathrm{int}$ quoted in Table~\ref{ephem}. The latter spin-down rate value was used to infer the pulsar's spin-down power $\dot E$, and magnetic field intensities at the stellar surface, $B_s$, and at the light cylinder, $B_\mathrm{LC}$.

Using the measured projected semimajor axis of the pulsar orbit, $x$, and the orbital period, $P_b$, we calculated the mass function given in Table~\ref{ephem}, as:

\begin{eqnarray}
f ( m_p, m_c ) = \dfrac{(m_c \sin i)^3}{(m_p + m_c)^2} = \dfrac{4 \pi}{T_\odot} \dfrac{x^3}{P_b^2},
\label{massfunc}
\end{eqnarray}

where $m_p$ is the pulsar mass, $m_c$ is the companion mass, $i$ is the inclination of the orbit, and $T_\odot = G M_\odot / c^3 = 4.925490947$~$\mu$s. Assuming a canonical mass of 1.4~M$_\odot$ and an edge-on orbit ($i = 90^\circ$), we obtain a lower limit on the companion mass of $\sim$$0.023$~M$_\odot$. Since the probability of observing a binary system with an inclination lower than $i_0$ for a random distribution of inclinations is given by $1 - \cos \left( i_0 \right)$, a 90\% confidence upper limit on the companion mass can be determined by assuming an inclination of 26$^\circ$. We find an upper limit on the companion mass of $\sim$$0.053$~M$_\odot$. As can be noted from Table~\ref{ephem}, the analysis of the NRT timing dataset revealed a significant $\dot P_b$ term, indicative of orbital period variations. These variations, which are commonly observed in other BW systems such as PSRs B1957+20, J1731$-$1847, or J2051$-$0827 \citep[see for instance][and references therein]{Bates2011,Shaifullah2016}, the very low companion mass and the observation of eclipses around superior conjunction (see Section~\ref{s:eclipses}) firmly establish PSR~J2055+3829 as a BW pulsar.

\subsection{Gamma-ray analysis}
\label{s:gamma}

The pulsar's spin-down power value $\dot E \sim$ $4 \times 10^{33}$ erg s$^{-1}$ is above the empirical deathline for gamma-ray emission from MSPs \citep{Guillemot2016}, and is comparable to that of many gamma-ray-detected MSPs \citep[see for example][]{Fermi2PC,Smith2019}. PSR~J2055+3829 is thus a candidate for a detection in GeV gamma rays with the \textit{Fermi}-LAT \citep{LAT}. No gamma-ray counterpart is found in the recently released \textit{Fermi}-LAT 8-year Source Catalog \citep[4FGL, see][]{4FGL} within 2$^\circ$ of PSR~J2055+3829. However, since pulsation searches are more sensitive than searches for continuous gamma-ray emission \citep[see][for recent examples of gamma-ray pulsars not reported in LAT source catalogs]{Smith2019}, we searched the LAT data for gamma-ray pulsations from PSR~J2055+3829. We analyzed $\sim$$10.2$ years of Pass 8 SOURCE class \textit{Fermi}-LAT events, with energies from 0.1 to 100 GeV, found within 3$^\circ$ of the pulsar's sky position, and with zenith angles smaller than 105$^\circ$. Pulse phases were calculated using the \texttt{fermi} plugin of \textsc{Tempo2} \citep{Ray2011} and the timing solution described above. The sensitivity of our gamma-ray pulsation searches was enhanced using the photon weighting scheme described in \citet{Bruel2019}, and we followed the same strategy as described in \citet{Smith2019} for searching the value of the $\mu_w$ weighting parameter that maximizes the significance of gamma-ray pulsations; that is, we tested three values of $\mu_w$: 3.2, 3.6, and 4.0, on the full LAT dataset available (MJD 54682 until MJD 58421), and on the validity interval of the timing solution (MJD 57333 to MJD 58421). None of the six search trials resulted in a pulsation significance above 2$\sigma$. With its Galactic latitude of $-4.259^\circ$, PSR~J2055+3829 lies close to the Galactic plane and is also located near the gamma-ray-bright Cygnus region. Because of the LAT's relatively large point-spread function at low energies, the selected dataset was likely contaminated by a large number of background gamma-ray photons. We repeated the analysis described above using a smaller event selection radius of 1$^\circ$, and obtained consistent results.

The non-detection of gamma-ray pulsations may result from the fact that the current timing solution for PSR~J2055+3829 is not able to maintain phase-connection over the entire \textit{Fermi}-LAT dataset, thus preventing us from detecting faint pulsations. Unfavorable beaming geometry could also explain the absence of significant gamma-ray pulsations. \citet{Guillemot2014} for instance found marginal evidence for different viewing angle (the angle between the spin axis and the line of sight) distributions between gamma-ray-detected and undetected energetic and nearby MSPs, and postulated that the undetected ones are seen under small viewing angles. However, a more likely explanation for the non-detection resides in the pulsar's large distance compared to those of known gamma-ray MSPs. The DM distance estimated with the YMW16 model of $\sim$$4.6$ kpc, if close to the actual value, is indeed significantly larger than the typical distance of gamma-ray MSPs, of $\sim$1 kpc \citep[see e.g.,][]{Fermi2PC}. The lack of a counterpart in 4FGL within 2$^\circ$ of PSR~J2055+3829 means that the integrated energy flux above 0.1 GeV for this pulsar is less than $4 \times 10^{-12}$ erg cm$^{-2}$ s$^{-1}$, obtained with the same method as the 3-year sensitivity map, Figure~16 of \citet{Fermi2PC}, but using the 8-year 4FGL model extrapolated to the $\sim$$10.2$ years of data used in this work. This energy flux limit is smaller than the fluxes of LAT-detected MSPs at Galactic latitudes within 5$^\circ$ of that of PSR~J2055+3829 (see 4FGL). Besides, assuming that the pulsar has an efficiency of conversion of spin-down power $\dot E$ into gamma-ray luminosity $L_\gamma$ of 100\% leads to an expected gamma-ray energy flux $h = \dot E / (4 \pi d^2)$ of $1.4 \times 10^{-12}$ erg cm$^{-2}$ s$^{-1}$ at the distance of 4.6 kpc, smaller than the energy flux limit quoted above. The detection of pulsed high-energy emission from PSR~J2055+3829 would thus require significantly more gamma-ray data, or a favorable (\textit{i.e.}, sharp) gamma-ray pulse profile allowing detection in spite of faint emission \citep[see][for discussion]{Hou2014}. 



%

\section{Eclipse properties}
\label{s:eclipses}

\begin{figure*}
\centering
\includegraphics[width=18cm]{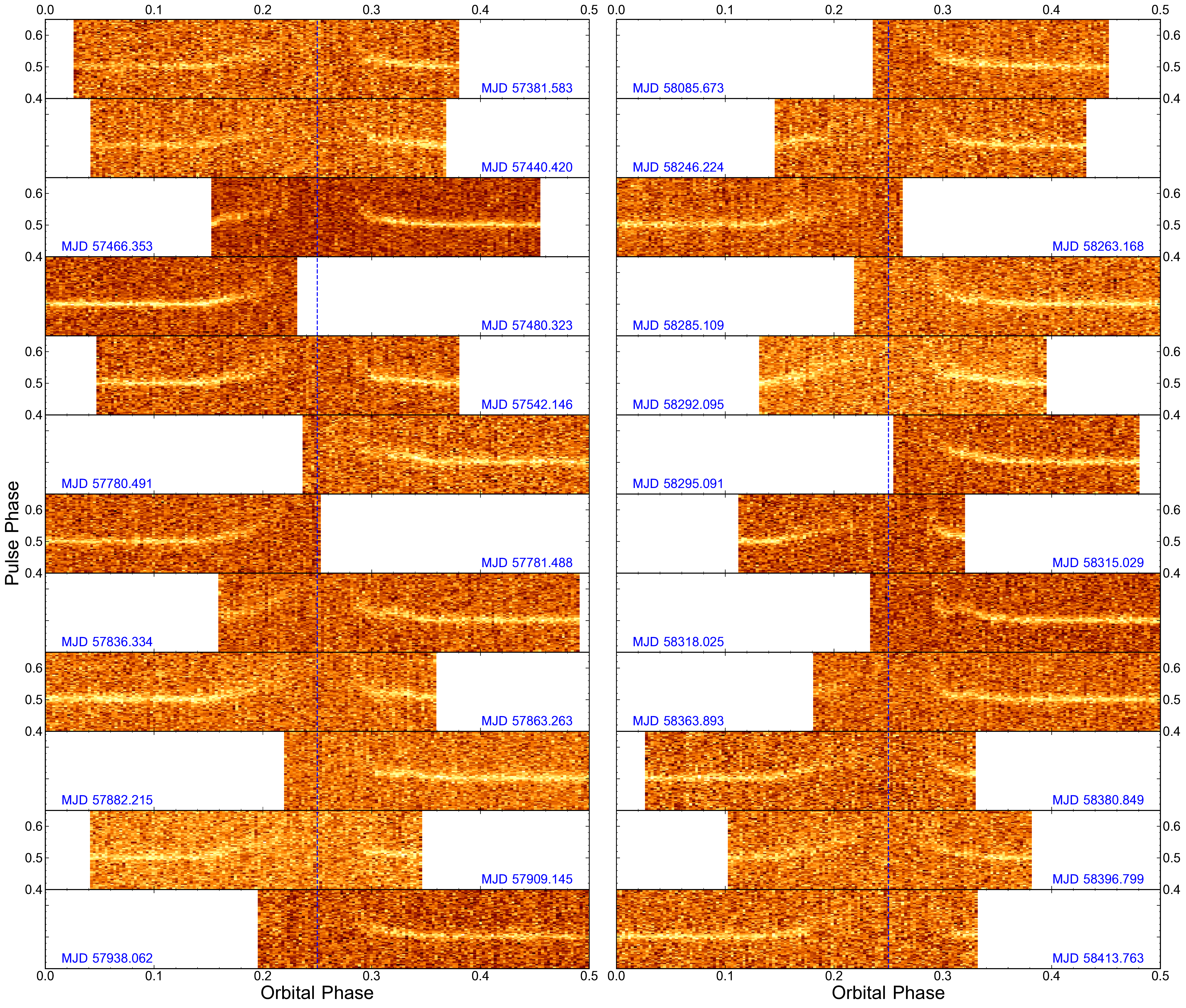}
\caption{Flux density as a function of orbital and pulse phase, for several observations of PSR~J2055+3829 at 1.4 GHz with the NUPPI backend. Observation start epochs are given in each panel. To improve readability, the number of phase bins was reduced by a factor of eight, and the time-resolution was decreased to form one profile per approximately 30 s. The dashed blue lines indicate the phase of superior conjunction (orbital phase 0.25).}
\label{eclipses}
\end{figure*}

As mentioned above, our 1.4 GHz observations of PSR~J2055+3829 revealed the presence of eclipses of the pulsar's radio signal around superior conjunction. A selection of high S/N detections of the pulsar containing complete or partial eclipse traverses is displayed in Figure~\ref{eclipses}. It is apparent from this Figure that the pulsar emission is completely obscured at 1.4 GHz for a few minutes around orbital phase 0.25. As is commonly observed in other BW systems displaying eclipses \citep[e.g.,][]{Bates2011,Bhattacharyya2013,Polzin2018}, we see asymmetric phase modulations in the pulsar signal at ingress and egress. The slightly longer egress phases suggest that the orbital motion of the companion causes its wind to be swept back \citep{Fruchter1990,Stappers2001}. The observations shown in Figure~\ref{eclipses} also indicate variations in the duration of the individual eclipses, as well as short-duration absorption events at egress (e.g. at MJD~58318) indicative of clumpiness in the outflow from the companion. More sensitive observations, or observations at lower radio frequencies would be useful for characterizing the eclipse-to-eclipse variability further.

For the timing analysis presented in Section~\ref{s:analysis}, we used a TOA dataset comprising one TOA per 10 min and per 128 MHz, with a total bandwidth of 512 MHz. With such a time- and frequency-resolution we could track long-term DM variations while keeping integration times short compared to the pulsar's orbital period. Nevertheless, in order to investigate short-term variations of the DM around superior conjunction with increased sensitivity and time-resolution, we used the wide-band template matching technique implemented in the \texttt{PulsePortraiture} software library\footnote{\url{https://github.com/pennucci/PulsePortraiture}} \citep{Pennucci2014} to extract one TOA per 5 min of observation, and for the entire frequency bandwidth. The two-dimensional template profile was constructed by summing 40 observations of PSR~J2055+3829 well outside of its eclipses, keeping the full frequency resolution available. The results of the timing analysis with this new TOA dataset are given in Table~\ref{ephem} (see `2D template' column), and the corresponding timing residuals are plotted in Figure~\ref{res_time}. With the notable exception of the EFAC parameter of $\sim$$1.73$, indicative of TOA uncertainties that were likely underestimated, the parameters from this new timing analysis are consistent with those obtained with the standard TOA extraction technique presented in Section~\ref{s:analysis}.

\begin{figure}
\centering
\includegraphics[width=0.95\columnwidth]{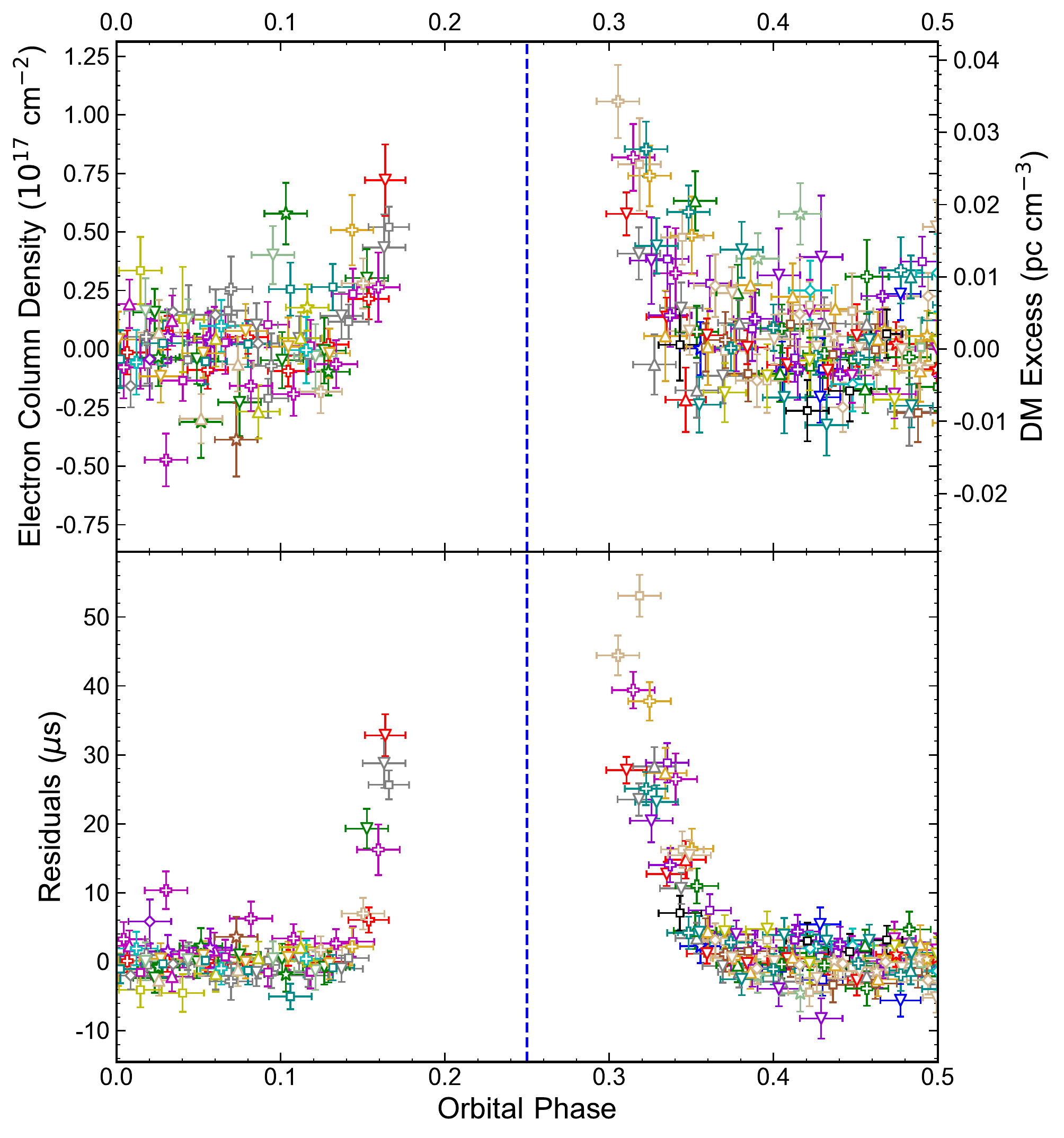}
\caption{Timing residuals and excess electron column density as a function of orbital phase, around superior conjunction. 
All data points in this figure correspond to 1.4 GHz observations, with S/N ratios larger than 19, close to the median value. Residuals and electron column densities from a given observation are displayed with the same combination of symbols and colors. The phase of superior conjunction is shown as a dashed blue line.}
\label{res_ophase}
\end{figure}

In addition to extracting the TOAs, we used the wide-band template matching technique to measure the DM of the pulsar in each 5-min data sample. The timing residuals and the DM values from this analysis around the eclipse phases are shown in Figure~\ref{res_ophase}. DM values plotted in this Figure were corrected for 
the long-term time variation measured from the timing analysis, and corresponding to the linear model plotted in Figure~\ref{DM_vs_time}. The DM offsets are converted to excess electron column densities, assuming that the material causing the delays is a homogeneous plasma. The eclipses are centered at orbital phase $\sim$$0.24$ and have a duration of about 19 min, or $\sim$$10$\% of the orbital period, where the center phase and duration quoted here simply correspond to the middle and separation of the last TOA before and first TOA after the eclipse. The asymmetry of the ingress and egress phases mentioned earlier is apparent in the timing residuals. To a good approximation, the radius of the companion's Roche lobe can be estimated to be \citep{Eggleton1983}:

\begin{eqnarray}
R_L = \dfrac{0.49 a q^{2/3}}{0.6 q^{2/3} + \ln \left( 1 + q^{1/3} \right)}, 
\end{eqnarray}

where $R_L$ the radius of the Roche lobe, $q = m_c / m_p$ is the companion and pulsar mass ratio, and $a$ is the separation between the companion star and the pulsar.   Assuming $m_p = 1.4$ M$_\odot$ and the minimum companion mass $m_\mathrm{c, min} = 0.02290$ M$_\odot$, we find $a \sim$$1.2$ R$_\odot$, and $R_L \sim$$0.14$ R$_\odot$. The eclipse duration translates to an opaque fraction of the companion's orbit of $\sim$$0.75$ R$_\odot$, which is larger than the companion's Roche lobe radius. This indicates that the eclipsing material is not gravitationally bound to the companion. At 1.4 GHz, the maximum added electron density near superior conjunction is found to be $N_{e,\mathrm{max}} \sim$$10^{17}$ cm$^{-2}$.


%

Table 4 of \citet{Bates2011} lists a number of useful eclipse properties for the extensively-studied eclipsing BW pulsars in the Galactic disk J1731$-$1847, B1957+20 and J2051$-$0827. Similar information on the eclipsing BWs PSR J1544+4937 and J1810+1744 can be found in \citet{Bhattacharyya2013} and \citet{Polzin2018}, respectively, although the latter articles did not report any orbital period derivative measurement. Comparing the values of the eclipse radius, Roche lobe radius, $P_b$, $| \dot P_b | / P_b$, $\dot E / a^2$, and $N_{e,\mathrm{max}}$ for PSR~J2055+3829 with those of other eclipsing BW pulsars, we find that the new MSP is similar to the previously-reported ones in most respects. The eclipse radius, Roche lobe radius, and maximum excess electron density of PSR~J2055+3829 are comparable to those of other eclipsing BWs. For the new MSP we find $\dot E / a^2 \sim 4.6 \times 10^{32}$ erg s$^{-1}$ lt-s$^{-2}$. Although this value for the $\dot E / a^2$ parameter is the lowest among those of the pulsars in this category, it is very close to the value for PSR~J2051$-$0827 of $4.8 \times 10^{32}$ erg s$^{-1}$ lt-s$^{-2}$, and is still several orders of magnitude higher than the median $\dot E / a^2$ value for other pulsars in binary systems. As was noted by \citet{Bates2011}, this strong energy flux at the distance of the companion could explain why ablation occurs in BW systems and not in other binaries. Finally, we find that our value of $| \dot P_b | / P_b$ is an order of magnitude lower than those of PSRs J1731$-$1847, B1957+20, and J2051$-$0827. Although statistically significant, the $\dot P_b$ value given in Table~\ref{ephem} should be taken with a grain of salt since the $\dot P_b$ values of BWs are known to vary rapidly \citep[see e.g. Figure~5 of][for the case of PSR~J2051$-$0827]{Shaifullah2016}. The many similarities PSR~J2055+3829 shares with other known eclipsing BWs suggest that they were formed by a common process. Observations of this new MSP thus provide a new insight into the origin of the population of BWs, and into eclipse mechanisms in these objects.

\section{The mass functions of eclipsing and non-eclipsing black widows}
\label{s:massfuncs}

\citet{Freire2005} compared the properties of eclipsing binary pulsars in the Galactic disk and in globular clusters (GCs), and found evidence for eclipsing BWs having higher mass functions than non-eclipsing ones. This can be understood as follows: eclipsing systems are thought to be seen under higher inclinations than non-eclipsing ones. Since the mass function is proportional to $\left( \sin i \right)^3$ (see Equation~\ref{massfunc}), and assuming that the pulsar and companion masses in these systems are comparable to those in eclipsing systems, non-eclipsing BWs are thus expected to have lower mass functions.

\begin{table}
\caption[]{Orbital properties of Galactic disk and globular cluster black widow pulsars. For each pulsar (see Section~\ref{s:massfuncs} for details on the selection criteria) we list the orbital period $P_b$, the projected semi-major axis $x$, and the mass function $f$. The last column indicates whether the pulsar is known to exhibit radio eclipses near superior conjunction.}
\label{ecl_table}
\centering
\small
\begin{tabular}{ccccc}

\hline
\hline
Pulsar & $P_b$ (days) & $x$ (lt-s) & $f$ ($10^{-6}$ M$_\odot$) & Eclipses? \\
\hline
\multicolumn{5}{c}{Galactic disk black widow pulsars} \\
\hline
J0023+0923 & 0.139 & 0.035 & 2.36 & -- \\
J0251+2606 & 0.204 & 0.066 & 7.42 & Y \\
J0610$-$2100 & 0.286 & 0.073 & 5.21 & -- \\
J0636+5129 & 0.067 & 0.009 & 0.18 & -- \\
J0952$-$0607 & 0.267 & 0.063 & 3.69 & -- \\
J1124$-$3653 & 0.227 & 0.080 & 10.67 & Y \\
J1301+0833 & 0.272 & 0.078 & 6.89 & Y \\
J1311$-$3430 & 0.065 & 0.011 & 0.30 & Y \\
J1446$-$4701 & 0.278 & 0.064 & 3.65 & -- \\
J1513$-$2550 & 0.179 & 0.041 & 2.31 & Y \\
J1544+4937 & 0.121 & 0.033 & 2.61 & Y \\
J1731$-$1847 & 0.311 & 0.120 & 19.24 & Y \\
J1745+1017 & 0.730 & 0.088 & 1.38 & -- \\
J1805+0615 & 0.338 & 0.088 & 6.42 & -- \\
J1810+1744 & 0.148 & 0.095 & 42.03 & Y \\
J1946$-$5403 & 0.130 & 0.043 & 5.23 & -- \\
B1957+20 & 0.382 & 0.089 & 5.23 & Y \\
J2017$-$1614 & 0.098 & 0.043 & 8.89 & Y \\
J2051$-$0827 & 0.099 & 0.045 & 10.01 & Y \\
J2052+1219 & 0.115 & 0.061 & 18.43 & Y \\
J2055+3829 & 0.130 & 0.045 & 5.93 & Y \\
J2115+5448 & 0.135 & 0.044 & 5.02 & Y \\
J2214+3000 & 0.417 & 0.059 & 1.28 & -- \\
J2241$-$5236 & 0.146 & 0.026 & 0.87 & -- \\
J2256$-$1024 & 0.210 & 0.081 & 12.94 & Y \\
\hline
\multicolumn{5}{c}{Globular cluster black widow pulsars} \\
\hline
J0024$-$7204I & 0.230 & 0.038 & 1.16 & -- \\
J0023$-$7203J & 0.121 & 0.040 & 4.86 & Y \\
J0024$-$7204O & 0.136 & 0.045 & 5.35 & Y \\
J0024$-$7204P & 0.147 & 0.038 & 2.72 & -- \\
J0024$-$7204R & 0.066 & 0.034 & 9.31 & Y \\
J1518+0204C & 0.087 & 0.057 & 26.82 & Y \\
J1641+3627E & 0.117 & 0.037 & 3.97 & Y \\
J1701$-$3006E & 0.159 & 0.070 & 14.76 & Y \\
J1701$-$3006F & 0.205 & 0.057 & 4.78 & -- \\
J1748$-$2446O & 0.260 & 0.112 & 22.40 & Y \\
J1807$-$2459A & 0.071 & 0.012 & 0.39 & -- \\
J1824$-$2452G & 0.105 & 0.017 & 0.44 & -- \\
J1824$-$2452J & 0.097 & 0.025 & 1.77 & -- \\
J1824$-$2452L & 0.226 & 0.057 & 3.90 & -- \\
J1836$-$2354A & 0.203 & 0.046 & 2.61 & -- \\
J1911+0102A & 0.141 & 0.038 & 2.88 & -- \\
J1953+1846A & 0.177 & 0.078 & 16.44 & Y \\
\hline

\end{tabular}
\end{table}

\begin{figure*}
\centering
\includegraphics[width=19cm]{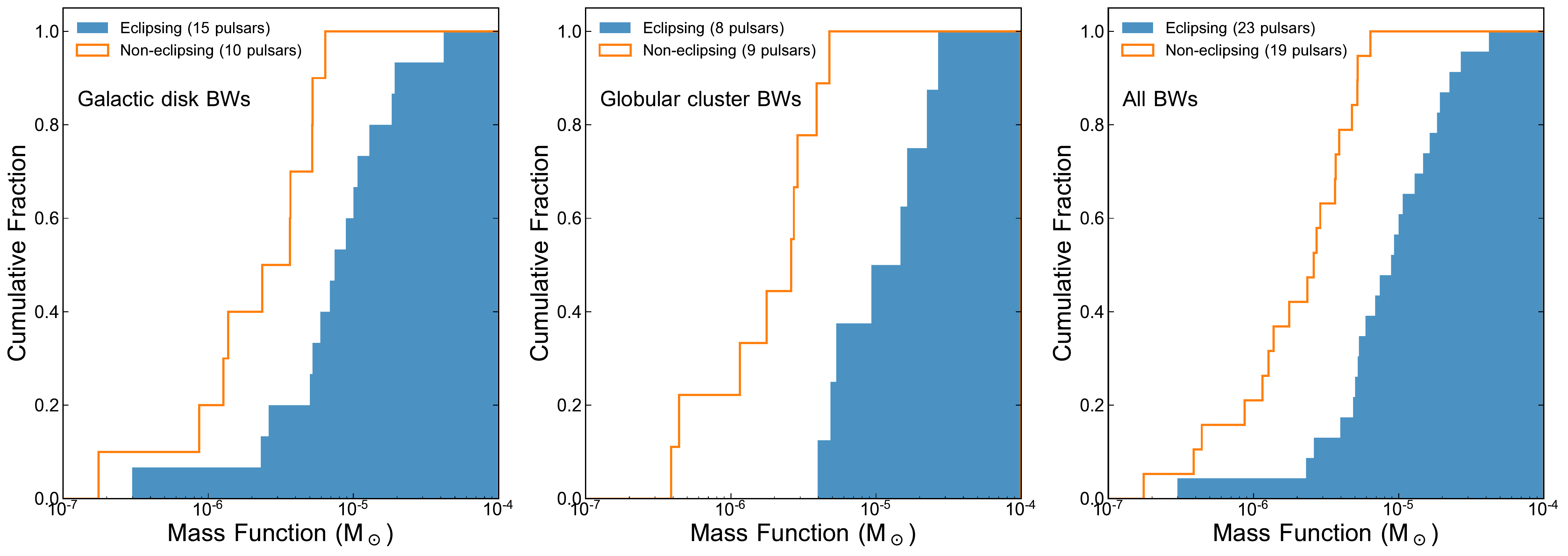}
\caption{Cumulative histograms of the mass functions (see Equation~\ref{massfunc}) of known eclipsing and non-eclipsing pulsars in the Galactic disk (left panel), in globular clusters (middle panel) and for the total population of BW pulsars (right panel). Pulsar names and mass functions for the various populations are given in Table~\ref{ecl_table}.}
\label{ecl_noecl}
\end{figure*}

Motivated by the fact that the number of known BW pulsars in the Galactic disk and in GCs has increased substantially since the \citet{Freire2005} study was published, we revisited the mass function distributions of eclipsing and non-eclipsing BWs. We selected Galactic disk\footnote{A list of Galactic disk MSPs is available at \url{http://astro.phys.wvu.edu/GalacticMSPs/GalacticMSPs.txt} .} and GC\footnote{See \url{http://www.naic.edu/~pfreire/GCpsr.html} for a list of known GC pulsars and their main properties.} BW MSPs (here defined as MSPs orbiting objects with masses $m_c < 0.06$ M$_\odot$, and not known to be planets), with published information regarding the presence of radio eclipses or lack thereof. Table~\ref{ecl_table} lists the selected objects, and some of their main orbital properties. The orbital parameters quoted in Table~\ref{ecl_table} were taken from the Australian Telescope National Facility (ATNF) pulsar database\footnote{\url{http://www.atnf.csiro.au/people/pulsar/psrcat/}} \citep{ATNF} when available, otherwise they were taken from the public lists of Galactic disk and GC MSPs. Cumulative histograms of the mass functions of eclipsing and non-eclipsing BWs in the Galactic disk, in GCs, and of the total population of known BWs are shown in Figure~\ref{ecl_noecl}. One-dimensional Kolmogorov-Smirnov (KS) tests \citep{Press1992} indicate that the probability that eclipsing and non-eclipsing objects originate from the same parent distribution is about 1.5\% in the case of Galactic disk MSPs, only $\sim 0.08$\% for GC BWs, and $\sim 0.007$\% for the total population of BWs. Therefore, the mass function distributions of eclipsing and non-eclipsing BWs do appear to be drawn from distinct distributions in all three cases, with eclipsing objects generally having higher mass functions than others. Two caveats we must point out are that not all MSPs considered as non-eclipsing have been observed at multiple radio frequencies (and in particular at low frequencies) so that some of them may eventually be found to exhibit eclipses, and the fact that non-eclipsing pulsars seen under low orbital inclinations may have heavier companions than actual BW pulsars. Despite these important caveats, this study gives credence to the hypothesis that eclipsing BWs have higher inclination angles (and thus higher mass functions) than non-eclipsing BWs. Simulations of populations of BW systems seen under many inclination angles and using realistic companion and pulsar mass distributions are beyond the scope of this paper, but may provide an avenue for investigating the observed differences further.

\begin{figure}
\centering
\includegraphics[width=0.95\columnwidth]{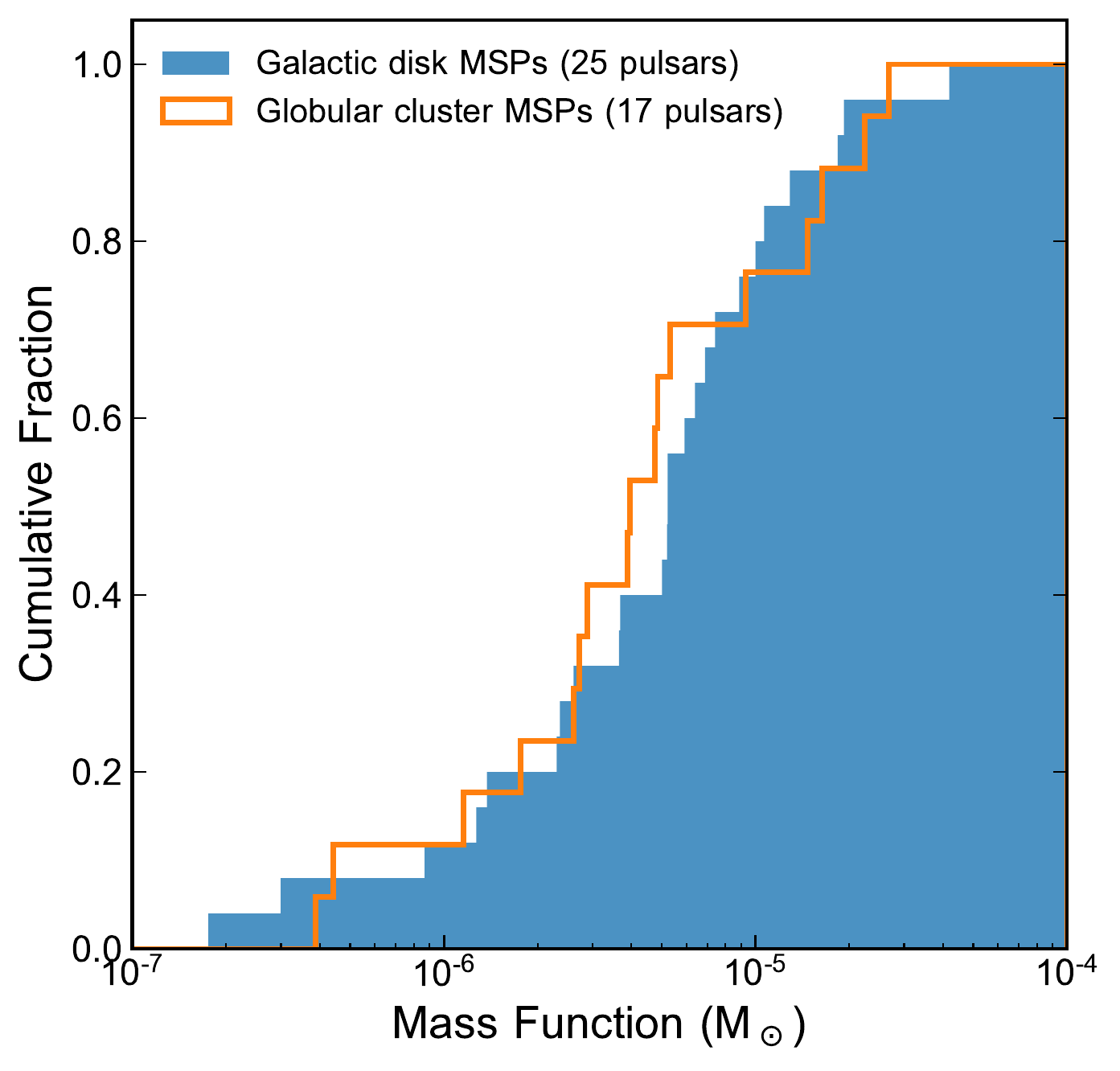}
\caption{Cumulative histograms of the mass functions of BW pulsars in the Galactic disk and in globular clusters.}
\label{gc_vs_fields}
\end{figure}

In addition to comparing the mass function distributions of eclipsing and non-eclipsing BWs, we also compared the mass functions of Galactic disk and GC BW pulsars (\textit{i.e.}, considering eclipsing and non-eclipsing ones), finding no obvious differences between their distributions. A comparison of the distributions (plotted in Figure~\ref{gc_vs_fields}) with the KS test gives a probability that they originate from a common parent distribution of about 50\%, that is, Galactic disk and GC BWs appear to have consistent mass function distributions. This indicates strong similarities between the last stages of the evolution of BWs in the Galactic disk and in globular clusters, despite the very different conditions and interactions in these environments, particularly the fact that MSPs in globular clusters evolved from X-ray binaries formed by exchange interactions, unlike Galactic disk MSP systems that evolve from primordial binaries \citep[see e.g.][]{Freire2013}. 


%

\section{Summary}
\label{s:summary}

We report timing of an MSP in a BW system, PSR~J2055+3829, originally discovered as part of the SPAN512 pulsar survey conducted at the Nan\c{c}ay Radio Telescope. A number of these BW pulsars are known to exhibit complex orbital instabilities \citep[see e.g.][and references therein]{Shaifullah2016} and are thus not stable enough for PTA applications, although we note that \citet{Bochenek2015} presented arguments in favor of their inclusion in PTAs. Continued timing observations of PSR~J2055+3829, whose apparent rotation appears to be relatively stable so far, will enable us to evaluate its long-term stability. Additionally, the long-term monitoring of this new pulsar may eventually make it possible to detect gamma-ray pulsations, although the low expected energy flux at the distance of 4.6 kpc makes the detection of PSR~J2055+3829 challenging, as discussed in Section~\ref{s:gamma}.

At 1.4 GHz, the pulsar is observed to be eclipsing for about 10\% of the orbit. As is also seen in other similar systems, the eclipses of PSR~J2055+3829 are asymmetric and variable. We also find that a number of its eclipse properties (e.g., the eclipse radius, the maximum excess column density near superior conjunction, or the $\dot E / a^2$ flux parameter) resemble those of previously studied BW systems. More generally, we find significant differences between the mass function distributions of eclipsing and non-eclipsing BWs in the Galactic disk, in GCs, and for the total population of known BWs. Eclipsing BWs tend to have higher mass functions than non-eclipsing ones, possibly because they are seen under higher inclination angles. On the other hand, Galactic disk and GC BWs have consistent mass function distributions, indicating that similar late phases of the evolutionary processes are at play in these different environments.

As mentioned in Section~\ref{s:eclipses}, more sensitive observations of PSR~J2055+3829 or observations at lower frequencies, where the eclipses and the dispersion of the radio signal by the outflow from the companion are more pronounced, would be particularly useful for further characterizing the nature of the eclipses and the interactions between the pulsar flux and the plasma released by the companion star. For instance, by observing the original BW pulsar B1957+20 with the 305-m Arecibo telescope at 0.3 GHz, \citet{Main2018} were able to detect extreme plasma lensing events near superior conjunction of the pulsar, making it possible to detect the pulsar during individual rotations and thus resolve its magnetosphere. Observations of plasma lensing events in B1957+20 were also used for probing the magnetic field at the interface between the pulsar and companion winds \citep{Li2019}. Another recent example is the low-frequency observations of the BW pulsar J1810+1744 with LOFAR and Westerbork presented in \citet{Polzin2018}, which enabled fine analyses of DM, flux density and scattering time variations during eclipse traverses. Finally, we note that optical observations of BW systems can provide information on the heating of the companion, and can be used to constrain the pulsar masses \citep[e.g.][]{Breton2013}. X-ray observations of BW pulsars can also be used to study the emission arising from the intrabinary shock of the pulsar wind \citep[see for instance][]{Gentile2014}, providing useful insight into the properties of this wind. No X-ray or optical observations of this pulsar have yet been conducted. New observations of PSR~J2055+3829 at complementary wavelengths and energies are therefore warranted.

\begin{acknowledgements}

The Nan\c{c}ay Radio Observatory is operated by the Paris Observatory, associated with the French Centre National de la Recherche Scientifique (CNRS). We acknowledge financial support from the ``Programme National de Cosmologie and Galaxies'' (PNCG), ``Programme National Hautes Energies'' (PNHE), and ``Programme National Gravitation, R\'ef\'erences, Astronomie, M\'etrologie'' (PNGRAM) of CNRS/INSU, France. PCCF gratefully acknowledges financial support by the European Research Council for the ERC Starting grant BEACON and continued support from the Max Planck Society under contract No. 279702. We gratefully acknowledge support from the CNRS/IN2P3 Computing Center (CC-IN2P3---Lyon/Villeurbanne) for providing the computing resources needed for the pulsar searches in the SPAN512 survey data.

\newline
The \textit{Fermi} LAT Collaboration acknowledges generous ongoing support from a number of agencies and institutes that have supported both the development and the operation of the LAT as well as scientific data analysis. These include the National Aeronautics and Space Administration and the Department of Energy in the United States, the Commissariat \`a l'Energie Atomique and the Centre National de la Recherche Scientifique / Institut National de Physique Nucl\'eaire et de Physique des Particules in France, the Agenzia Spaziale Italiana and the Istituto Nazionale di Fisica Nucleare in Italy, the Ministry of Education, Culture, Sports, Science and Technology (MEXT), High Energy Accelerator Research Organization (KEK) and Japan Aerospace Exploration Agency (JAXA) in Japan, and the K.~A.~Wallenberg Foundation, the Swedish Research Council and the Swedish National Space Board in Sweden.
 
Additional support for science analysis during the operations phase is gratefully acknowledged from the Istituto Nazionale di Astrofisica in Italy and the Centre National d'\'Etudes Spatiales in France. This work performed in part under DOE Contract DE-AC02-76SF00515.

\newline
We would like to thank Pablo Saz Parkinson for his helpful comments and suggestions. 

\end{acknowledgements}

\bibliographystyle{aa} 
\bibliography{2055}

\end{document}